\documentstyle[12pt,aaspp4,natbib,psfig]{article}
\newcommand{\simgt}{\lower 2pt \hbox{$\, \buildrel {\scriptstyle >}\over {\scriptstyle\sim}\,$}}
\newcommand{\simlt}{\lower 2pt \hbox{$\, \buildrel {\scriptstyle <}\over {\scriptstyle\sim}\,$}}

\def\asca{{\it ASCA\/}}

\def\einstein{{\it Einstein\/}}

\def\heao1{{\it HEAO-1\/}}
\def\hst{{\it {\it HST}\/}}

\def\iue{{\it IUE\/}}
\def\rosat{{\it ROSAT\/}}

\newcommand{\pgfif}{PG~1535+547}
\newcommand{\pgfour}{PG~1411+442}
\newcommand{\pgten}{PG~1011--040}
\newcommand{\pgbal}{PG~2112+059}
\newcommand{\pgoh}{PG~0043+008}
\newcommand{\pgsev}{PG~1700+518}
\newcommand{\aox}{$\alpha_{\rm ox}$}

\lefthead{Gallagher, Brandt et al.}
\righthead{X-ray Absorption in Soft X-ray Weak AGN}
\begin{document}
%

\title{Heavy X-ray Absorption in Soft X-ray Weak Active Galactic Nuclei}
\author{S. C. Gallagher,\altaffilmark{1} 
W. N. Brandt,\altaffilmark{1}
A. Laor,\altaffilmark{2}
M. Elvis,\altaffilmark{3} 
S. Mathur,\altaffilmark{4}
Beverley J. Wills,\altaffilmark{5} 
and N.~Iyomoto\altaffilmark{6}}
\altaffiltext{1}{Department of Astronomy and Astrophysics, The Pennsylvania State University,
525 Davey Lab, University Park, PA 16802, USA} 
\altaffiltext{2}{Physics Department, Technion, Haifa 32000, Israel}
\altaffiltext{3}{Harvard-Smithsonian Center for Astrophysics, 60 Garden Street, 
Cambridge, MA 02138, USA} 
\altaffiltext{4}{Department of Astronomy, Ohio State University, 174 West 18th Avenue, 
Columbus, OH 43210-1106, USA}
\altaffiltext{5}{Department of Astronomy, University of Texas at Austin, Austin, 
TX 78712, USA}
\altaffiltext{6}{The Institute of Space and Astronautical Science,
3-1-1 Yoshinodai, Sagamihara, Kanagawa 229-8510, Japan}
%
\setcounter{footnote}{0}

\begin{abstract}
\noindent
Recent \rosat\ studies have identified a significant population 
of Active Galactic Nuclei (AGN) that are notably faint in soft X-rays relative
to their optical fluxes.  Are these AGN intrinsically X-ray weak or are 
they just highly absorbed?
Brandt, Laor \& Wills have systematically
examined the optical and UV spectral properties of a well-defined sample of 
these soft X-ray weak (SXW) AGN drawn from the Boroson \& Green sample of 
all the Palomar Green AGN with $z<0.5$.  
We present \asca\ observations of three of these SXW AGN: \pgten, \pgfif\ (Mrk~486), and \pgbal.
In general, our \asca\ observations support the intrinsic absorption scenario for 
explaining soft X-ray weakness; both \pgfif\ and \pgbal\ show significant
column densities ($N_{\rm H}\approx10^{22}$--$10^{23}$~cm$^{-2}$)
of absorbing gas.  Interestingly, \pgten\ shows no spectral evidence for
X-ray absorption. The weak X-ray emission may result from very strong absorption 
of a partially covered source, or this AGN may be intrinsically X-ray weak.
\pgbal\ is a Broad Absorption Line (BAL) QSO, and we find it to have the 
highest X-ray flux known of this class.  It shows a typical power-law X-ray continuum
above 3~keV; this is the first direct evidence that BAL QSOs indeed have normal X-ray
continua underlying their intrinsic absorption.
Finally, marked variability between the \rosat\ and \asca\ observations of \pgfif\ and
\pgbal\ suggests that the soft X-ray weak designation may be transient, and multi-epoch
0.1--10.0~keV X-ray observations are required to constrain variability of the absorber
and continuum.
\end{abstract}



\section{Introduction}

The canonical, blue X-ray loud AGN have received much attention though
there are other numerically significant AGN populations \citep[e.g.,][]{El1992}.  
Here we study one of the `minority' populations: UV-excess AGN ($U-B\le-0.44$)
with weak X-ray emission.
X-ray weak AGN were first noted in \einstein\ data \citep[e.g.,][]{ElFa1984,AvTa1986}.
\rosat\ extended these studies and identified a significant population of type~1 
AGN notably faint in 
the soft X-ray band (0.1--2.0~keV) relative to their optical 
fluxes \citep[e.g.,][]{LaEtal1997}.  
Although not as strong as the radio-loud/radio-quiet dichotomy \citep[e.g.,][]{KeEtal1989},
soft X-ray weakness is quite a strong effect. 
The Laor et al. (1997) soft X-ray weak (SXW) AGN lie at least an order of magnitude
below the mean of the X-ray to optical flux ratio for the Palomar Green (PG) 
AGN sample \citep{ScGr1983}.
Possible explanations for this
distinct trait are intrinsic X-ray absorption, 
inherently different spectral energy distributions, 
and variability of the X-ray and/or optical fluxes.  
Only significant optical variability has been disallowed 
by optical monitoring \citep[e.g.,][]{GiEtal1999}.
SXW AGN have yet to be systematically investigated with X-ray
spectroscopy in order to determine the cause for their faintness.

Brandt, Laor \& Wills (2000; hereafter BLW) \nocite{BrLaWi2000} 
have identified and studied the SXW 
AGN population in the Boroson \& Green (1992; hereafter BG92) \nocite{BoGr1992} 
sample of all PG~QSOs with $z<0.5$.
Specifically, they took SXW~AGN to be those with the optical to X-ray 
spectral index, 
$\alpha_{\rm ox}\leq-2$.{\footnote {$\alpha_{\rm ox}$ is
the slope of a power law defined by the flux densities at
rest-frame 3000 \AA\ and 2 keV. A large negative value of 
$\alpha_{\rm ox}$ indicates weak X-ray emission.  
The mean value for radio-quiet 
AGN is $\approx -1.48$ \citep[e.g.,][]{LaEtal1997} 
with a typical range from $-1.7$ to $-1.3$ (e.g., BLW).  Other authors 
prefer $\alpha^{\prime}_{\rm ox}$, the slope of a power law between
the flux densities at rest-frame 2500 \AA\ and 2 keV.  For a slope of 
$\alpha_{\rm u}$ between 2500 and 3000 \AA, 
$\alpha^{\prime}_{\rm ox}$=1.03\aox$-0.03\alpha_{\rm u}$ (BLW).
Values of $\alpha^{\prime}_{\rm ox}$ from the literature have been converted to \aox\ 
(assuming $\alpha_{\rm u}=-0.5$) to facilitate direct comparisons; the original values
are indicated in parentheses.}}
In soft X-rays, these sources are 10--30 times fainter 
relative to their optical fluxes than most AGN.
This effect is much more extreme than any possible luminosity dependence of
\aox\ \citep[e.g.,][]{GrEtal1995}.  For example, even the most luminous objects in 
Green et al. (1995) have \aox~$\ge-1.7$.

BLW demonstrated a strong correlation
between $\alpha_{\rm ox}$ and the equivalent width (EW) of intrinsic 
C~{\sc iv} absorption, suggestive of a 
physical link between soft X-ray weakness and UV absorption.
Intrinsic X-ray absorption does depress
soft X-ray flux; however, the underlying X-ray power-law continuum will
recover in the harder 2--10~keV band if $N_{\rm H}\simlt5\times10^{23}$~cm$^{-2}$.  
The absorption hypothesis is thus 
testable with spectroscopic X-ray observations which makes such sources 
ideal targets for \asca\/. 
\asca\ offers access to penetrating X-rays and thus the ability 
to probe much larger column densities of absorbing gas than \rosat\/.
If soft X-ray weakness does
indeed arise from intrinsic absorption, this is a significant
step towards understanding the link between X-ray and UV absorbers. 
Precisely relating the X-ray and UV absorbers has proved difficult in
practice although there is highly suggestive evidence for some
connection (e.g., Mathur et al. 1998, BLW, and references therein). \nocite{MaWiEl1998}

We have undertaken a project to observe the 10 SXW~AGN 
studied by BLW over a broad X-ray band to test the absorption interpretation 
for a significant number of objects.  
In addition to identifying absorption, X-ray spectral analysis can constrain the nature of
the absorbing gas, e.g., column density, ionization state, covering factor, and velocity
distribution. These quantities are important for
determining the total mass outflow rates from AGN.  
We proposed \asca\ observations of the seven SXW~AGN from BLW that had not yet been
observed, and here we report on the three we were able to observe 
before the end of the normal \asca\ operations phase.
The SXW AGN in this study are more extreme in soft X-ray weakness 
than warm absorber AGN,
and they cover a large range in UV absorption up to the strength of Broad Absorption Lines (BALs).
BAL~QSOs lie at the extreme end of the BLW correlation of 
C~{\sc iv} absorption EW with \aox, but those
examined thus far (e.g., \pgoh\ and \pgsev; Gallagher et al. 1999)
have not offered enough X-ray photons for significant spectroscopic work.  
Therefore, X-ray analyses of BAL QSOs
have always assumed an underlying X-ray power law the same as that observed in normal QSOs.
We aim to investigate by direct spectroscopic observations 
whether the range of absorption properties represents a 
continuum in the X-rays.  
If this is the case, it would support the assumption that the 
extreme X-ray weakness observed in BAL~QSOs is due to heavy intrinsic absorption.

Basic properties of the observed SXW AGN and the
observations are listed in Table~1.  Following, we present background
information on each of the objects.

\noindent 
{\bf \pgten}:
The small H${\beta}$ Full Width at Half Maximum (FWHM) of 1440~km~s$^{-1}$
for this SXW AGN makes it a Narrow-Line Seyfert~1;  
its type~1 nature is supported by the presence of Fe~{\sc ii} 
emission (BG92).  \pgten\ does not show any evidence for
 C~{\sc iv} or Ly$\alpha$ absorption lines
in analyses of available \iue\ spectra (BLW)
and is potentially useful for comparison with the other targets 
for exploring whether UV and X-ray absorption are always related.

\noindent
{\bf \pgfif~(Mrk~486, 1 ~Zw~1535+55)}:
This source has strong optical Fe~{\sc ii} emission first noted by de Veny \& Lynds 
(1969)\nocite{DeLy1969}, 
and it is the X-ray weakest, detected SXW~AGN in the BLW sample.  
It has a Narrow-Line Seyfert~1 spectrum with H${\beta}$ FWHM=1480~km~s$^{-1}$
and is notable as the PG~AGN with the highest optical continuum polarization 
\citep[$P$=2.5\%;][]{BeScWeSt1990}. 
The structure in the polarization and C~{\sc iv} absorption 
has been studied in detail by Smith et al. (1997).
\pgfif\ was the only radio-quiet AGN observed by Neugebauer \& Matthews (1998) 
to exhibit significant near-infrared variability from 1980--1998.
\nocite{SmScAlHi1997,NeMa1999}

\noindent
{\bf \pgbal}:
This SXW AGN is one of the most luminous low-redshift ($z<0.5$) PG~QSOs 
with $M_{\rm V}=-27.3$, and it
has the second largest C~{\sc iv} absorption EW, 19 \AA, of all the AGN in the BLW sample.
\hst\ spectra revealed broad, shallow C~{\sc iv} absorption making this
a BAL QSO with a balnicity index
of $\approx 2980$~km~s$^{-1}$ (Jannuzi et al. 1998; BLW; and see Figure~1), 
well above the lower limit that defines the class
(Weymann et al. 1991).\nocite{JaEtal1998,WeMoFoHe1991}
There are no spectra in the
literature that cover the Mg~{\sc ii} region, and so it is not known
whether \pgbal\ has broad Mg~{\sc ii} absorption lines (see Boroson \& Meyers 1992 for
further discussion of Mg~{\sc ii} BAL~QSOs) \nocite{BoMe1992}. This source was detected by
\rosat\ which is unusual for a BAL QSO \citep{KoTuEs1994,GrMa1996}.
Since few BAL~QSOs are even detected in X-rays, observations of 
this source provide a valuable test of the hypothesis that BAL QSOs have the same 
underlying X-ray continua as normal QSOs. 

\section{Observations and Data Analysis}

In Table~1, we list the relevant observation dates, exposure times, and instrument
modes for our targets, and SIS images for each source are displayed in Figure~2.  
For \pgfif\ and \pgbal, we were able to use serendipitous
X-ray sources in archival \rosat\ PSPC data to confirm the \asca\ pointing.  \pgten\
was only observed off-axis with \rosat; thus the \asca\ pointing could not be
independently confirmed.  However, 
\asca\ pointings are generally quite reliable, and the detected 
source position was coincident with the optical position within the expected error \citep{Go1996}.
The spectra resulting from these observations were extracted with {\sc xselect}, 
a program in the {\sc ftools} package, following the general procedures described in 
Brandt et~al. (1997b)\nocite{BrMaReEl1997}. We have used Revision~2 data 
\citep{Pi1997} and adopted the standard Revision~2 screening criteria.  In Table~2,
the \asca\ count rates or count rate upper limits for each detector are listed.
Count rates were determined by subtracting the counts in a source-free background 
region (normalized to the area of the source region) from those in the source 
region, and then dividing by the exposure time.  The count rates in Table~2 and 
the spectra below are filtered in 
energy from 0.6--9.5/0.9--9.5 keV in the SIS/GIS detectors in order to eliminate
poorly calibrated channels. Fluxes in the 0.5--2.0~keV and 2.0--10.0~keV bands 
are calculated from the best-fitting models with the SIS0 normalization
(see Table 2).  Unless otherwise noted, we used the rest-frame 
3000~\AA\ continuum flux densities from Neugebauer et al. (1987)
and the rest-frame 2~keV flux densities from the best-fitting SIS0 model to 
calculate \aox.
Galactic absorption is fixed in all of the model fits, and
errors stated are always for 90\% 
confidence conservatively taking all model parameters to be of interest other than 
absolute normalization.  
We adopt $H_0=70$~km~s$^{-1}$~Mpc$^{-1}$ and $q_0=\onehalf$ throughout.  

\subsection{\pgten}

\pgten\ was detected in all instruments except GIS2 at greater than 3$\sigma$ 
above the background.
Low signal-to-noise ratio spectra were successfully 
extracted from SIS0 and GIS3 for 
analysis with the X-ray spectral analysis tool {\sc xspec} \citep{Ar1996}. 
There were not enough counts in the SIS1 detector to warrant spectral analysis.
The background spectrum for SIS0 was 
extracted from a region on the CCD free from point source contamination, 
while that for GIS3 was extracted from blank-sky background files 
as detailed in Ikebe et al. (1995)\nocite{IkEtal1995} in order
to mitigate vignetting effects.  

The spectra were first fit with a power law with fixed Galactic
absorption.  The fit was statistically acceptable, 
and the resulting photon index, $\Gamma\approx2$ (see Table 3), is consistent with
those typically observed in radio-quiet type~1 AGN, $\Gamma=1.5$--2.5 
(Brandt et al. 1997a; Reeves \& Turner 2000).
\nocite{BrMaEl1997,ReTu2000}
However, \pgten\ has an extremely
low 2--10~keV X-ray luminosity, $L_{2-10}$, of $10^{41.8}$~erg~s$^{-1}$.
In order to determine if X-ray absorption is the cause of the weakness of \pgten\
in soft X-rays, we examined the spectrum for any indications of absorption.
If the intrinsic neutral column density were $\simlt10^{23}$~cm$^{-2}$, then the 
X-ray spectrum would start to recover to the level of the incident power law by 
$\sim3$--5~keV.  Such a recovery is not evident in the spectrum, and thus we do not 
see evidence for absorption of this magnitude (see Figure~3).
Given the current data, we constrain the intrinsic
column density to be $N_{\rm H}\le5\times10^{21}$~cm$^{-2}$,
though the poor photon statistics prevent detailed modeling of the data.  
Another possibility is that the observed power law may not be the direct nuclear source, 
but only an indirect scattered component.
If this is the case, then the direct component can reasonably be modeled as 
an additional power law with $\Gamma=2.0$ and the expected 1~keV normalization for an 
AGN of this optical brightness assuming an intrinsic \aox\ of $-1.5$. 
A large neutral column density of $N_{\rm H}>1.0\times10^{24}$~cm$^{-2}$ 
is then required to extinguish the primary continuum throughout the \asca\ bandpass.
If the observed X-ray emission is only scattered light, one might expect
a narrow iron~K$\alpha$ emission line; 
given the poor quality of the data, the EW for a line in the range from 6.4--6.97~keV
can only be constrained to be $<6.6$~keV.

A nearby source $2.^{\prime}5$ away in
the SIS1 and GIS detectors complicated the measurement of count rates. 
We searched the Palomar Optical Sky Survey for an optical counterpart to this 
nearby source, but
none was apparent.  The confusing source was quite faint in SIS1, and the lack of 
its detection in SIS0
can be understood as resulting from the different position of each detector with
respect to the optical axis of the telescope.
For GIS2, the count rate upper limit for \pgten\ in Table~2
is $3\sqrt{N_{\rm s}}$/$t$, where ${N_{\rm s}}$ 
is the number of counts in the source cell and $t$ is the exposure time.  
Counts from the nearby source mentioned above were excluded.

We have investigated potential variability between the 350~s \rosat\ All-Sky Survey
observation and our \asca\ exposure.  Although the constraints are not tight,
we do not find highly significant evidence for variability.
\subsection{\pgfif\ (Mrk~486)}
\pgfif\ was clearly detected in all four detectors with enough total counts, $\approx940$,
for spectral analysis.
The source spectra were extracted using circular source cells with 
radii $2.^{\prime}7$/$4.^{\prime}1$ for the SIS/GIS  
to maximize the signal-to-noise ratio, and the background regions were extracted as
described in $\S$2.1.
The following spectral analysis was also done using different
background regions on the CCDs for the SIS as well as source-free background 
regions on the GIS.
The results were consistent and thus are not sensitive to the choice 
of background region. 
 
For the first fit to the data in all four detectors, a power law with 
fixed Galactic absorption was used.  
The resulting photon index, $\Gamma=0.45^{+0.17}_{-0.22}$, 
is abnormally flat for a typical type~1 AGN.  In addition, the residuals showed a
systematic excess below 1~keV indicating that the fit was poor. 
A $\chi^2$ of 134.7 for 119 degrees of 
freedom was obtained indicating that this model can be rejected with 
84\% 
confidence according to the $P(\chi^2|\nu)$ probability distribution as defined in 
$\S 11.1$ of Bevington \& Robinson (1992; see Table~3). 
\nocite{BeRo1992} Next, we fit a power law with 
fixed Galactic absorption also allowing for the presence of intrinsic absorption.
The fitted photon index was again low, $\Gamma=0.44^{+0.25}_{-0.32}$, 
with negligible intrinsic absorption and
$\chi^2$=134.7 for 118 degrees of freedom; this model can be rejected with 86\% 
confidence.  A plot of the residuals again showed an
excess below 1~keV.  Forcing the photon index to lie in the
typical range for radio-quiet AGN, $\Gamma=$1.5--2.5,
only worsened the fit.  Finally, we fit a model adding 
intrinsic neutral absorption partially covering the incident power law.
The choice of this model was motivated by the presence of low energy photons which
would be completely absorbed by total coverage of the emission source
by neutral gas.
An effective partial covering situation could result from two lines-of-sight to the source of 
the X-ray emission.  In this scenario, the first component would be direct emission
absorbed by gas along the line-of-sight, and the second component would be unabsorbed
emission scattered by electrons towards the observer.  
Since scattering results in an increased fraction
of polarized flux,
the high observed optical polarization of \pgfif\ suggests that scattering is
indeed occurring in the nucleus.
For the partial covering model, the resulting best-fitting model parameters were an 
intrinsic neutral absorption column density of 
$(1.23^{+0.83}_{-0.54})\times10^{23}$~cm$^{-2}$, a covering fraction of 
$91^{+7}_{-26}$\%,
 and $\Gamma=2.02^{+0.92}_{-0.95}$  (see Figure~4).  
The $\chi^2$ was 117.5 for 117 degrees of freedom, a significant improvement
over all of the previous fits at the 99\% 
confidence level according to the $F$-test.  
The residuals showed no obvious trends,
and the partial covering model explains the additional soft photons 
evident in the simple absorption
models while also providing a physically reasonable photon index.  
Confidence contour plots for the two 
partial covering parameters are displayed in Figure~5.
From this best-fitting model to the \asca\ data, 
we obtain $L_{2-10}$=10$^{42.1}$~erg~s$^{-1}$ 
(absorption-corrected $L_{2-10}$=10$^{42.4}$~erg~s$^{-1}$).
For completeness, we also tried an ionized absorber model, but the fit
left clear systematic residuals.  The partial-covering absorption model was preferable,
with $\Delta\chi^2=-17$ for the same number of degrees of freedom.

According to the \asca\ data, the measured 
$\alpha_{\rm ox}=-2.03$ (see Table~2) 
is greater by $\Delta$\aox=0.42 than that calculated in BLW 
from the 2678~s \rosat\ PSPC observation performed on 1993 February 8.  
This dramatic change corresponds to
a factor of $\approx13.5$ increase in flux for a fixed X-ray spectral shape.
However, a closer look at the PSPC data indicates that \pgfif\ was not actually detected in the 
0.5--2.0~keV band but only in the 0.1--0.5~keV band.  
In the 0.5--2.0~keV band, there were 5 photons in the
$1.^{\prime}2$ radius source cell with 2 expected from the background.
Using the method of Kraft, Burrows \& Nousek (1991),
\nocite{KrBuNo1991} the upper limit on the number of counts from the target is 8.54 at the 95\%
confidence level.  This corresponds to an observed count rate of $<3.2\times10^{-3}$~ct~s$^{-1}$.
To estimate \aox, BLW assumed a 
photon index of $\Gamma=3.1$ based on the H$\beta$ FWHM (see $\S$3.3 of Laor et al. 1997
for discussion) and then found the normalization
that gave the proper count rate for the entire 0.1--2.5~keV band.  
If instead the 0.5--2.0~keV count rate upper limit 
is used to estimate an upper limit on the 2~keV flux density, then
\aox\ is measured to be $<-2.17$.  The lower limit on the 
flux increase for a fixed spectral shape is then a factor of 2.4. 
Though the limited photon statistics in the \rosat\ observation preclude spectral analysis,
we can still compare the two observations in order to determine if the increase
in \aox\ is greater than expected from measurement and systematic errors.
We took the best-fitting \asca\ SIS0 spectral model and convolved it with
the \rosat\ PSPC response using {\sc xspec}.  From this, we predicted 
a PSPC count rate of $(4.9^{+1.1}_{-0.9})\times10^{-3}$~ct~s$^{-1}$ in the
0.5--2.0 keV band (errors are for 90\% confidence), 
clearly above the measured upper limit.
Though the absolute calibrations of the two satellites have a systematic offset, 
measured fluxes in the 0.5--2.0 keV band are generally consistent within errors 
\citep{IwFaNa1999}; this cannot account for the increase in brightness.
The source has varied significantly in the six years between 
the two observations. 



\subsection{\pgbal}
In all four \asca\ instruments, \pgbal\ was clearly detected with enough counts,
a total of $\approx1950$, for spectral analysis.
Source spectra were extracted as described above using circular source cells with 
radii $2.^{\prime}6$/$3.^{\prime}8$ arcmin for the SIS/GIS. 
 The GIS source cell was chosen slightly smaller
than in $\S$2.2 in order to avoid including photons from a faint source
located $\approx4.^{\prime}1$  from the target.  Background spectra were also 
extracted as described in $\S$2.1, and we have verified that our results below
are not sensitive to the details of the background subtraction.

For the first fit to the data in all four detectors, a power law with fixed
Galactic absorption was used.  A $\chi^2$ of 182.8 for 151 degrees of 
freedom was obtained indicating that this model can be rejected with 96\%
confidence.
 The resulting photon index, 
$\Gamma=1.44^{+0.08}_{-0.09}$, 
is rather flat for a typical QSO.
In addition, the residuals showed a systematic deficit below 1~keV 
which also indicated that the fit was 
not good in the soft band and suggested the presence of intrinsic absorption.  
In order to model only the continuum, we excluded the data below 3~keV in the rest frame
(2 keV in the observed frame).  Above 3 keV, the absorption of incident photons by 
neutral or partially ionized gas should be much less than in the softer band, 
thus allowing the underlying power-law continuum to be observed and fit directly.
These data were then fit with 
a power law plus Galactic absorption.  This resulted in a steeper photon index of
$\Gamma=1.94^{+0.23}_{-0.21}$, more characteristic of radio-quiet QSOs, and 
$\chi^2$=69.0 for 76 degrees of freedom.  The data below 3~keV were then included,
and the model extrapolated to the lowest \asca\ energy bins (0.87 keV rest frame).  
The resulting plot, shown in Figure~6,
is suggestive of the possible nature of the intrinsic absorber.  The spectra
appear to recover towards the power-law continuum in the lowest energy bins, hinting at either 
partially ionized absorbing gas or partial covering of the continuum.
In contrast, a neutral, cold absorber
would depress almost all flux below $\approx3$~keV, the energy where the continuum recovers.  
We fit several models to try to characterize the absorbing gas.  The details 
of each fit are listed in Table~3.  
Briefly, intrinsic absorption by neutral gas with column
density $1.1\times10^{22}$~cm$^{-2}$ was preferable at the $>99$\% 
confidence level to a straight power law with only Galactic absorption
according to the $F$-test ($\Delta\chi^{2}=-40.6$).  
In addition, adding neutral intrinsic absorption
resulted in a photon index of $\Gamma=1.97^{+0.26}_{-0.23}$ which is consistent with $\Gamma$
for the fit to the data above 3~keV in the rest frame (see Figure~7).  
Fitting the data with a warm absorber or a partial covering model also 
yielded acceptable fits to the data; however, the fits were not statistically 
significant improvements over that with a neutral absorber (see Table 3).  
An iron K$\alpha$ line is not detected; we can set an upper limit on the EW of a narrow
line from 6.4--6.97~keV of $<210$~eV. 

\pgbal\ is a luminous X-ray source 
with $L_{1-2}\approx10^{43.7}$ and 
$L_{2-10}\approx10^{44.6}$~erg~s$^{-1}$, and so any possible 
starburst X-ray contribution would negligibly influence our spectral fitting.
Though absorption of one form or another is certainly required, these data do not have a high
enough signal-to-noise ratio to distinguish between various absorption models. 
Regardless of the nature of the absorber, the photon index is consistently $\approx2.0$, 
completely typical for a radio-quiet QSO.  
We have investigated the impact of the changing SIS calibration
on the results of our spectral-fitting (T. Yaqoob 1999, private communication) and
find that our general conclusions are not substantively affected.

The archival \rosat\ PSPC data from a 21.1 ks observation on 1991 November 15
cannot be simultaneously fit with the \asca\ data without 
allowing the power-law normalization to drop by a factor of $>7$.
If we convolve the \asca\ SIS0 model with the \rosat\ PSPC response, then the predicted 
0.5--2.0~keV count
rate for the \rosat\ observation is $(1.83^{+0.13}_{-0.16})\times10^{-2}$~ct~s$^{-1}$.
This value is almost a factor of four larger than the observed PSPC count rate in the same band
of $(4.8\pm0.7)\times10^{-3}$~ct~s$^{-1}$.
The \rosat\ data have a low signal-to-noise ratio, but spectral analysis
allows for a measurement of the 2~keV flux density which is relatively insensitive
to the specific model.  Recalculating \aox\ more robustly from spectral fitting 
of the data (rather than assuming a model and using the PSPC count rate as in BLW),
one obtains \aox$=-2.07$, similar to the BLW value of $-2.11$.  The measured
\aox$=-1.75$ from the best-fitting \asca\ data is significantly higher than any 
\aox\ consistent with the \rosat\ data; 
$\Delta$\aox\ of 0.32 between the two observations corresponds
to an increase in the rest-frame 2~keV flux density by a factor of $\approx7.2$.
The discrepancy between the relative increases in the 2~keV flux density and the observed 
versus predicted PSPC count rates can be reconciled if the X-ray spectral shape has varied, 
perhaps due to changes
in the absorbing material along the line-of-sight.  For example, a change in ionization state
of the material along the line-of-sight could strongly affect the rest-frame 2~keV flux density 
but cause smaller changes to the observed frame 0.5--2.0~keV integrated flux.
Clearly, \pgbal\ has brightened substantially in the soft X-ray band  
in the decade since the \rosat\ observation.  
In fact, \pgbal\ is no longer SXW as defined by BLW.  

\section{Discussion and Conclusions}

\subsection{\pgten}
From our rudimentary spectral analysis, \pgten\ appears to have a typical
power-law X-ray continuum with no evidence for absorption beyond the Galactic
column density.  
Forcing a normal underlying AGN X-ray spectrum in addition to the 
observed component requires an intrinsic neutral column
density of $N_{\rm H}>10^{24}$~cm$^{-2}$ to extinguish it to the top of the \asca\ bandpass; 
such a large column density would be extraordinary for a type~1 AGN.
Detecting a narrow iron~K$\alpha$ emission line with a large EW would provide evidence for
such a situation, but we are unable to set meaningful constraints with the 
current data. Given that the exact relationship between 
UV and X-ray absorbers is still unclear, such heavy absorption can perhaps be 
reconciled with the lack of detected C~{\sc iv} absorption if the 
gas is dense and cold or very close to the nucleus.  Dense, cold gas can effectively
absorb X-rays without creating strong absorption in the high-ionization C~{\sc iv} line
as can gas lying interior to the UV emission region.
Alternatively, this may be an intriguing example of an AGN with intrinsically 
weak X-ray emission. 
However, the normal He~{\sc ii}~$\lambda$4686 emission (EW$=10.2$ \AA; BG92) 
suggests that the intrinsic X-ray emission cannot
be extremely abnormal (Korista, Ferland \& Baldwin 1997; BLW). \nocite{KoFeBa1997}
A depressed ionizing continuum in the extreme UV (EUV) would not produce 
significant He~{\sc ii} emission, and it is difficult to reconcile a normal EUV continuum with 
drastically reduced X-ray emission.
Regardless, \pgten\ merits further study in both the
X-ray and UV regions of the spectrum in order to more tightly constrain the apparent 
lack of absorption in both bands (see Figures~8 and 9).  
In addition, UV emission line studies of this object are important for further investigating
the intrinsically weak X-ray continuum scenario.

The \asca\ data provide a measurement of $\alpha_{\rm ox}$ 
consistent with that derived from the \rosat\ PSPC count rate, and so this 
source remains firmly in the SXW category.  
From archival \einstein\ data, $\alpha_{\rm ox}=-1.89$ ($\alpha^\prime_{\rm ox}=-1.93$) 
was measured \citep{TaEtal1986}, indicating that this source has been
consistently weak in soft X-rays for at least two decades.  \iue\ data from 
1993 and 1995 show no evidence for UV variability.

\subsection{\pgfif\ (Mrk~486)}
\pgfif\ shows clear signs of intrinsic absorption by a large column density
of gas ($N_{\rm H}\approx1.2\times10^{23}$~cm$^{-2}$)
with only partial covering of the power-law continuum.  
The partial covering model is particularly interesting 
with regard to the optical polarization properties of this AGN since it 
implies possible electron scattering of X-rays around intrinsic absorbing clouds.
This result is consistent with the known high polarization of the optical continuum flux
($P$=2--8\%; Smith et al. 1997)
and closely resembles the X-ray results for the SXW AGN \pgfour\ \citep{BrWaMaYu1999}.  
\pgfour\ has a comparable internal column density, $N_{\rm H}$=$1.3\times 10^{23}$ cm$^{-2}$,
and covering fraction, $f_{\rm cov}$=$0.97^{+0.02}_{-0.09}$.  
From the broad-band spectral energy distribution of \pgfif\ (see Figure~8), 
the reddened continuum in the optical and UV is clearly seen, as well 
as the recovery of the X-ray continuum to within the range expected given the 3000 \AA\
flux density.  The amount of reddening in the optical and UV, though significant, 
is still much less than the $A_{\rm V}\approx76$
expected along the direct line-of-sight if the absorbing material has roughly
a Galactic dust-to-gas ratio \citep{BuHe1978}.  
Though dust may be responsible for the optical polarization, 
it is an inefficient scattering medium for X-rays; electrons are  
the most likely candidates for X-ray scatterers.

The increase in X-ray brightness of \pgfif\ in the six years between the \rosat\ and \asca\ 
observations is notable.  Determining whether the continuum flux has increased or
the absorption has decreased is not possible given the limited \rosat\ data.  During 
the same epoch, the near-infrared flux of \pgfif\ in 4 bands from 1.27--3.7~$\mu$m 
also varied significantly, increasing by $\approx0.5$ mag from
1993 to 1996 and then dropping by $\approx0.5$ mag between 1996 and 1998 \citep{NeMa1999}.
Our analysis of archival UV data from \iue\ indicates that on
1982 November 17 and 1984 May 7, \pgfif\ was 30\% 
brighter than in earlier \iue\ spectra (1982 June 6 and September 26).  Later \hst\ spectra
(1992 September 19) are consistent with the earliest \iue\ data.
\subsection{\pgbal}
With this \asca\ observation, \pgbal\ is found to have the largest X-ray flux of
any BAL~QSO known.
For comparison, the next brightest \asca-observed BAL~QSO, PHL 5200, had broad-band
count rates $\approx3$ times lower in all detectors \citep{MaElSi1995,GaEtal1999}.
Our result of a typical QSO X-ray continuum with absorption is robust. Regardless of 
the nature of the absorbing gas, neutral, partially ionized or partially covering,
the photon index is $\approx2.0$.
This is the first direct spectral evidence for a typical QSO X-ray continuum in a BAL QSO 
and supports the assumption that they are indeed normal QSOs cloaked by absorbing gas
\citep[e.g.,][]{GrMa1996,GaEtal1999}.  Our best-fitting model to the \asca\ data
finds an intrinsic, neutral 
column density of $\approx10^{22}$~cm$^{-2}$.  The data are not of sufficient quality to 
determine the ionization state of the gas; ionized gas would require a larger column 
density for the same amount of continuum absorption.  An additional issue 
complicating the determination of the line-of-sight column density is the unknown
velocity dispersion of the gas.  Currently available
models for X-ray absorption by neutral or partially ionized gas do not usually account 
for any velocity dispersion within the absorber.  Thus they are severely limited for
the study of BAL~QSOs which may have velocity dispersions as large as $\sim10^4$~km~s$^{-1}$
if the X-ray and UV absorbing gas have similar velocity structure.  At these high
dispersions, bound-bound X-ray absorption becomes significant and may dominate
over bound-free absorption (see $\S2.2$ of Netzer 1996; D. Chelouche \& H. Netzer 2000, 
private communication).
\nocite{Ne1996}

The measured neutral column density for \pgbal\ is more than an order of magnitude lower 
than the intrinsic column densities implied for the BAL~QSOs PG~0946+301 \citep{MaEtal2000}
 and \pgsev\ (Gallagher et al. 1999), and it may be that
this lower column density is related to the relatively shallow BAL troughs of this 
object (see Figure~1).
Future observations of a larger sample of BAL~QSOs can test if the depths in BAL troughs
are correlated with the observed X-ray column density, but care must be taken to evaluate the 
amount of scattering which can partially `fill in' BALs.  
The measured column density also
appears to be insufficient for the `hitchhiking' gas postulated by Murray et al. (1995).
\nocite{MuChGrVo1995}  In their model, thick, highly ionized gas with 
$N_{\rm H}\approx10^{23}$~cm$^{-2}$ 
absorbs soft X-rays and shields the BAL wind from becoming completely stripped of electrons.  
The X-ray absorbing gas is stalled at the base of the
flow and is distinct from the gas causing the broad UV absorption lines. 
Though our data suggest that this is not the case for \pgbal,  
higher spectral resolution observations are required to constrain the velocity 
of the X-ray absorbing gas.

From \einstein\ data, Tananbaum et al. (1986) measured an upper limit of \aox~$<-1.78$ 
($\alpha^{\prime}_{\rm ox}<-1.82$) which is weaker than the current value, but consistent
with the \rosat\ observations.
Since then, the variability of $\alpha_{\rm ox}$ from $-2.07$ to $-1.75$
over the course of a decade makes further
UV studies as well as higher signal-to-noise ratio X-ray observations of this 
source important.  In particular, the demonstrated
variability of \pgbal\ in the X-rays can be compared with the UV BALs.  
The probable X-ray spectral variability hints at 
changes in absorption structure, either of the ionization parameter or the column density.
BAL QSOs generally have little variability in their UV absorption line profiles
\citep[e.g.,][]{Barlow1993}.  However, if the UV and X-ray absorbing gas
are closely related and the increase in X-ray flux is a result of changing absorption structure, 
then the dramatic change in $\alpha_{\rm ox}$ should occur in concert with significant changes 
in the UV absorption lines.  The gravitationally lensed QSO PG~1115+080 is an
example of a mini-BAL QSO which appears to show variable X-ray absorption \citep{Ch2000}
as well as clear changes in UV absorption (Michalitsianos, Oliversen \& Nichols 1996). 
\nocite{MiOlNi1996} 
The archival \hst\ FOS data of \pgbal\ were taken in 1992, approximately one year after the 1991 
\rosat\ observation and seven years before this \asca\ observation.
\pgbal\ is scheduled for a possible Cycle 9 \hst\ observation as part of a UV absorption 
snapshot survey.  

From the spectral energy distribution of \pgbal\ (see Figure~8), it is apparent
that the \hst\ UV spectra are not continuous
with the Neugebauer et al. (1987) UV photometric data points which we 
used to calculate \aox\ in Table~2.  In
the $\approx13$ years between the 1980 N87 measurements and the \hst\ FOS observation, 
it is likely that the UV/optical flux has diminished.  Optical variability is not 
unreasonable as between 1992 and 1996, part of the time between the 
\rosat\ and \asca\ observations, \pgbal\ brightened by 0.18 mag in the 
$V$ band \citep{RaMcSmSt1998}.  Additional photometric data up to the present epoch
show continued 
variability of $\pm0.07$ mag in $V$ (Garcia-Rissmann et al., in preparation).
Analysis of \iue\ and \hst\ data from 1986 to 1995 shows no evidence for 
UV continuum variability.
If instead of using N87 values the \hst\ continuum data are extrapolated to rest-frame 
3000 \AA, the \rosat\ and \asca\
values for \aox\ are $-1.94$ and $-1.62$, respectively. Our \asca\ absorption-corrected
\aox\ is then $-1.55$, within $1\sigma$ of 
the mean of the Laor et al. (1997) radio-quiet AGN distribution of \aox.  
This is further evidence supporting the
picture of BAL~QSOs as AGN with typical underlying X-ray continua.

Since the strongest signatures of X-ray absorption are edges 
at fairly low energies (\simlt~2~keV), the characteristics of the X-ray 
absorbing gas are most readily studied in low-redshift objects where rest-frame
photons below $\approx2$~keV can still be detected with available X-ray instruments.
With its high X-ray luminosity, $L_{1-2}\approx10^{43.7}$~erg~s$^{-1}$
(absorption-corrected $L_{1-2}\approx10^{44.2}$~erg~s$^{-1}$) and
$L_{2-10}\approx10^{44.6}$~erg~s$^{-1}$, \pgbal\ is unlikely
to suffer significant contamination by a nuclear starburst as the most X-ray luminous
starburst galaxy known is more than 200 times fainter \citep{MoLeHe1999}.
Two other relatively X-ray bright BAL QSOs, CSO~755 ($z=2.88$) and PHL~5200 ($z=1.98$), are
at much higher redshifts, and thus \pgbal\ offers the most promise for detailed 
absorption studies with the new generation of X-ray observatories. 
With higher spectral resolution and greater sensitivity at low energies, 
parameters such as the ionization state and column density of the absorber
can be probed more thoroughly.
Ideally, grating observations of this source have 
the potential to elucidate the dynamical structure of the X-ray absorbing gas. 

\subsection{X-ray Absorption as the Primary Cause of Soft X-ray Weakness}

The number of SXW AGN in BLW with 
published \asca\ spectral constraints is now six out of ten: 
PG~0043+008, \pgten, \pgfour, \pgfif, PG~1700+518, and \pgbal. 
As the number of SXW AGN observed in the 2--10~keV band grows, 
the general identification of soft X-ray weakness with intrinsic X-ray absorption becomes 
more plausible.
\pgfour\ had been shown previously to have substantial X-ray absorption 
with a neutral column density of $N_{\rm H}\approx1.3\times10^{23}$~cm$^{-2}$ 
\citep{BrWaMaYu1999}.
With two of our sample, \pgfif\ and \pgbal, we have also examined
objects with more X-ray absorption than is typically seen in type~1 AGN but still with enough
photons for direct spectroscopic analysis.
The BAL QSOs PG~0043+008 and PG~1700+518 were not detected by \asca\ which suggests 
that they suffer absorption by a column density larger than 
$\approx5\times10^{23}$~cm$^{-2}$ \citep{GaEtal1999}.
\pgten\ is the only object to show no evidence for strong UV or X-ray absorption.

With the exception of perhaps \pgten, the SXW AGN which have been spectroscopically 
studied in the hard band show no evidence for unusual intrinsic spectral energy distributions.
In addition, the occurrence of UV and X-ray absorption is now one-to-one; we have found
evidence for X-ray absorption in those sources with detected UV absorption.
These observations lend direct support to absorption as the primary cause for
soft X-ray weakness as put forward by BLW.  After correcting the observed X-ray spectra of
\pgfour, \pgfif, and \pgbal\ for intrinsic absorption, one obtains photon indices and 
\aox\ values consistent with typical type~1 AGN. This serves to emphasize the 
universality of X-ray emission in QSOs, and the value of $\alpha_{\rm ox}$ as a 
parameter for identifying QSOs with interesting UV and X-ray absorption properties.

The three AGN in this study and \pgfour\ are ideal targets for higher throughput
X-ray spectroscopy and additional UV observations.  The most compelling issues to address
remain the nature of the relationship between the X-ray and UV absorbers and the 
dynamics of the X-ray absorbing gas.  Since the bulk of the gas in the inner
regions may be most readily observable in X-rays, it is essential to 
measure the velocity of this gas to determine the mass-outflow rates of AGN.

\subsection{X-ray and UV Variability}

The \aox\ variability of two SXW AGN, \pgfif\ and \pgbal, suggests that other sources
in the intermediate range ($-2.0<\alpha_{\rm ox}<-1.7$) may also vary and perhaps even join
the ranks of SXW AGN at times.  One possible example of such interesting variability 
is PG 0844+349 which is generally not SXW, but in one out of five X-ray observations 
had an intermediate \aox\ of $-1.88$ ($\alpha^{\prime}_{\rm ox}=-1.92$; Wang et al. 2000).
\nocite{WaEtal2000}

Additional 0.1--10.0~keV X-ray observations of the sources known to vary in 
$\alpha_{\rm ox}$ are essential for determining whether the
absorbing gas or the underlying continuum itself is changing. \pgbal\ appears to be an 
example of a SXW AGN whose spectral shape has changed over time most likely
as a result of changing absorption structure (see $\S$2.3).  Simultaneous or 
near-simultaneous UV observations are also required to elucidate the precise relationship
of the UV and X-ray absorbers and calculate \aox\ consistently.  
Currently, we have direct evidence for variability in
X-ray flux which translates into horizontal motion in the BLW C~{\sc iv} absorption EW 
versus \aox\ plot (see Figure~9).  
As of yet, we do not have corresponding multi-epoch C~{\sc iv} absorption EW
measurements and cannot determine if the plot is as dynamic vertically.  
Though monitoring of UV BALs by Barlow (1993) showed little variability, 
the matter requires further investigation.  If the variability is
a result of bulk motions of absorbing clouds across the line-of-sight, then the AGN 
currently towards the absorbed end of the correlation may show more dramatic 
changes over time as their clouds continue to move.

\subsection{Soft X-ray Weak AGN and the X-ray Background}

All three of the SXW~AGN studied above, and indeed all of the SXW~AGN of 
BLW, were originally selected by their blue optical color, and BLW found no
evidence for a systematic difference between the optical continuum
slopes of SXW and non-SXW AGN. Thus, given our X-ray spectral fitting
results, it appears that a non-negligible fraction of blue type~1 
AGN suffer significant X-ray absorption and have hard X-ray spectra 
qualitatively similar to that of the X-ray background. Such objects
need to be remembered when constructing AGN synthesis models for the 
X-ray background, and indeed some blue QSOs with hard X-ray spectra
have been found in X-ray background surveys 
\citep[e.g.,][]{FiEtal1999,BrEtal2000}.

\acknowledgements

We thank F. Hamann for thoughtful and constructive comments.
We acknowledge the support of a NASA Graduate Student Researchers Program Grant and
the Pennsylvania Space Grant Consortium (SCG), NASA LTSA grant NAG5-8107 
and NASA grant NAG5-7256 (WNB), and NASA LTSA grant NAG5-3431 (BJW).  
This research has made use of data obtained through the NASA Extragalactic Database (NED)
and the High-Energy Astrophysics Science Archive Research Center (HEASARC) Online Service, 
provided by NASA's Goddard Space Flight Center.





\begin{figure}[t!]
\hbox{
\psfig{figure=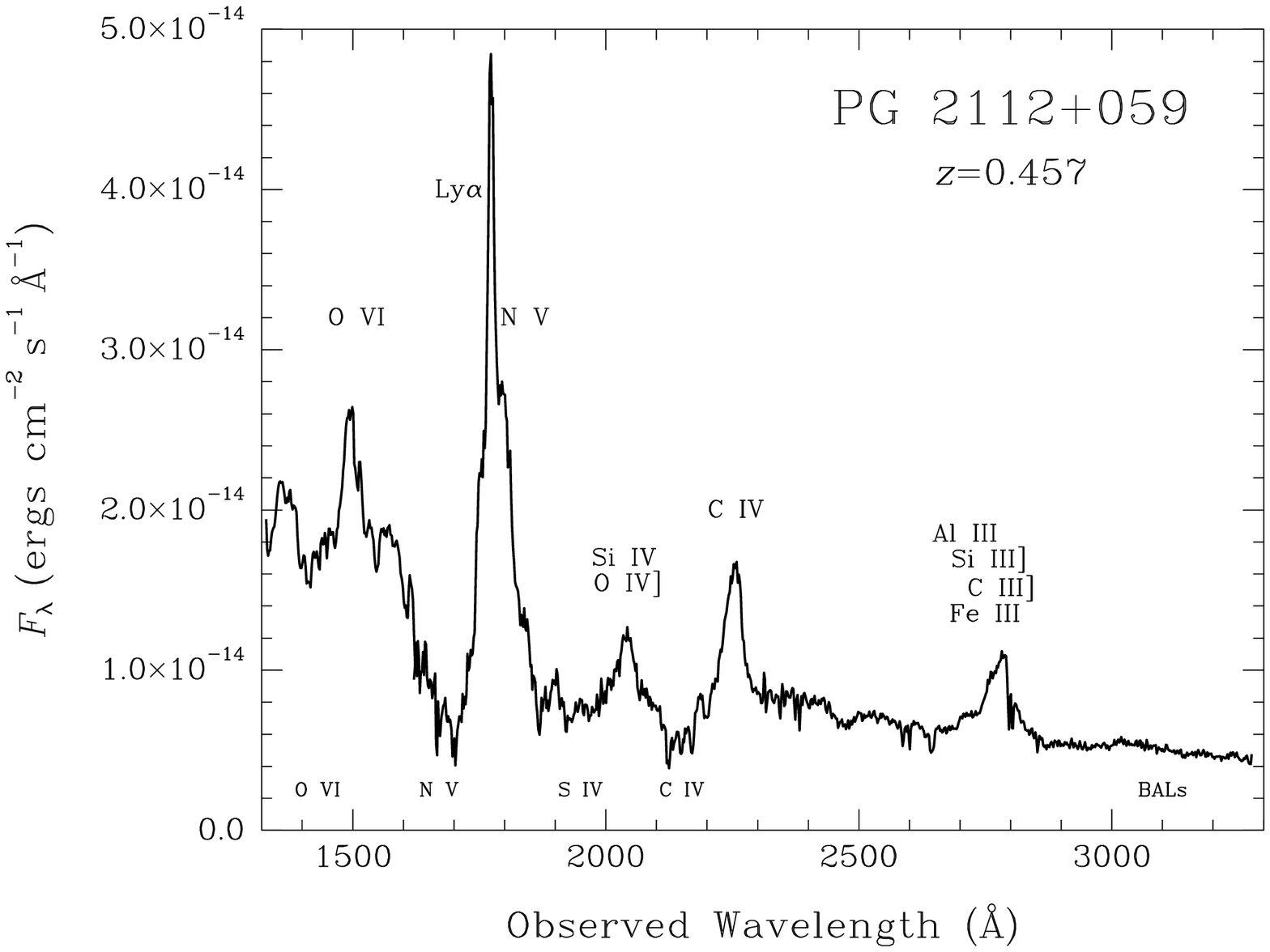,height=4.0truein,width=6.6truein,angle=0.}
}
\caption{
\hst\ UV spectrum of \pgbal.  The data from 1617--3277 \AA\ (observed frame) 
have been previously published
(Jannuzi et al. 1998), but the shorter wavelength data are presented here for the first time.
Observed wavelength ranges, observing modes, and observation dates are as follows: 
1327--1613 \AA, 
GHRS G140 grating, 1995 July 31; 1598--2311 \AA, FOS G190 grating, 1992 September
19; and 2223--3277 \AA, FOS G270, 1992 September 19.  The data have been 
dereddened using the Galactic $N_{\rm H}$ given in Table~1 and 
$E(B-V)=N_{\rm H}/(5.0\times10^{21} {\rm cm}^{-1})$ 
(Diplas \& Savage 1994;  Predehl \& Schmitt 1995).
The GHRS and FOS data are in excellent agreement in the region of wavelength
overlap even though the flux-density calibration may be uncertain for these small slit 
observations.  The combined data as presented here show more clearly than previously 
published spectra the shallow and very broad absorption lines against the QSO continuum 
interpolated between, e.g., 1600 and 2300 \AA.  There may be up to four equally spaced 
line-locked absorption systems identifiable in the O{\sc VI}$\lambda\lambda$1031,1037 and 
C{\sc IV}$\lambda\lambda$1548,1550 doublets and possibly some identified in Ly$\alpha$ and 
N{\sc V}$\lambda\lambda$1238,1242 (see Foltz et al. 1987 for an excellent example of 
line-locking in a BAL~QSO).
}
\end{figure}
\nocite{DiSa1994,PrSc1995,FoWeMoTu1987}
\clearpage
\begin{figure}[t!]
\hbox{
\psfig{figure=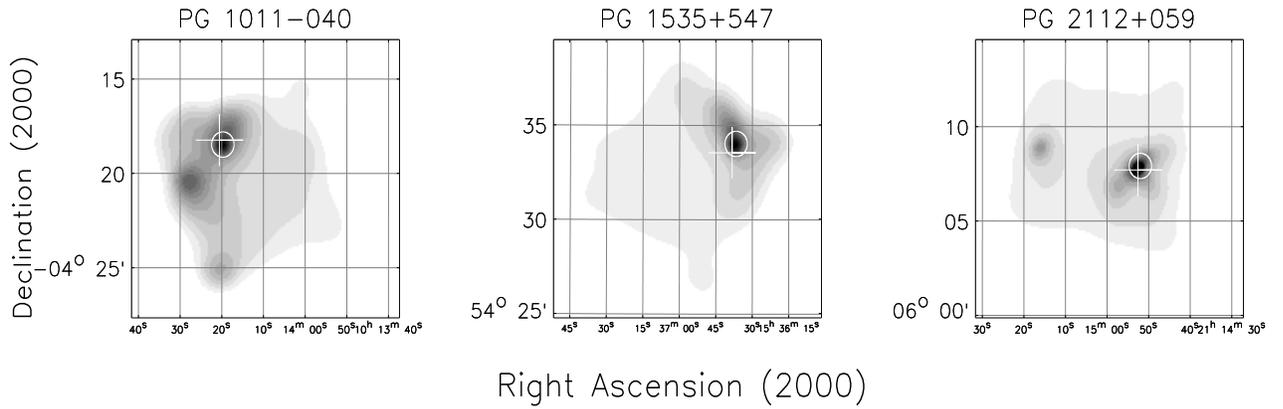,height=2.1truein,width=6.6truein,angle=0.}
}
\caption{
Full-band (0.6--9.5 keV) co-added SIS0 and SIS1 images for each of the SXW AGN in this study.
Images have been adaptively smoothed at the 3$\sigma$ level using the
algorithm of Ebeling, White \& Rangarajan (2000).   
The white cross marks the precise optical position of each source.
The white circle is centered on the X-ray source with a radius of $40^{\prime \prime}$
representing the standard \asca\ pointing error.
}
\end{figure}
\nocite{EbWhRa2000}
\clearpage
\begin{figure}[t!]
\hbox{
\psfig{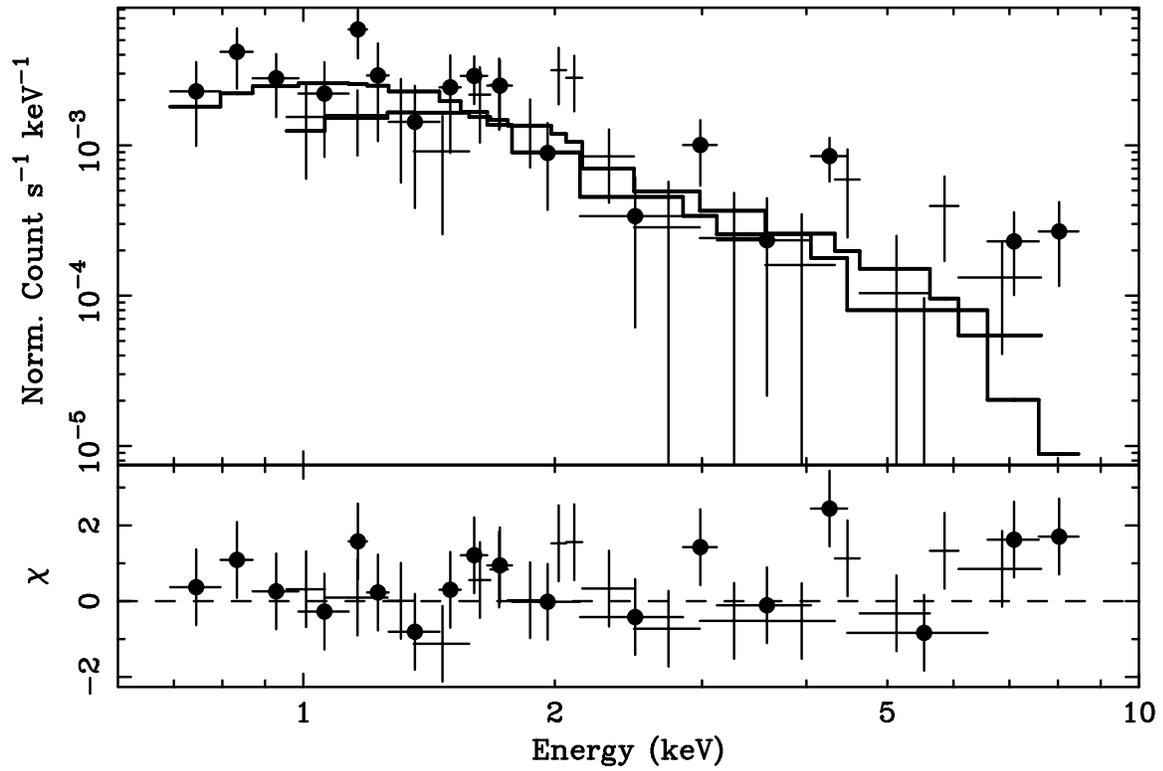}
}
\caption{\asca\ SIS0 and GIS3 observed-frame spectra of \pgten\ fit with a power-law
model. 
Filled circles are the data points for the SIS0 detector while plain crosses are
for the GIS3 detector. 
The ordinate for the lower panel, labeled $\chi$, shows the fit 
residuals in terms of $\sigma$ with error bars of size one.
}
\end{figure}
\clearpage
\newpage
\begin{figure}[t!]
\vspace{0.3in}
\hbox{
\psfig{figure=figure4.ps,height=4.0truein,width=6truein,angle=-90}
}
\caption{\asca\ SIS and GIS observed-frame spectra with the best-fitting model for \pgfif\ 
(see $\S$2.2). 
Filled circles are the data points for the SIS detectors while plain crosses are
for the GIS detectors. 
The ordinate for the lower panel, labeled $\chi$, shows the fit 
residuals in terms of $\sigma$ with error bars of size one.
}
\end{figure}
\clearpage
\begin{figure}[t!]
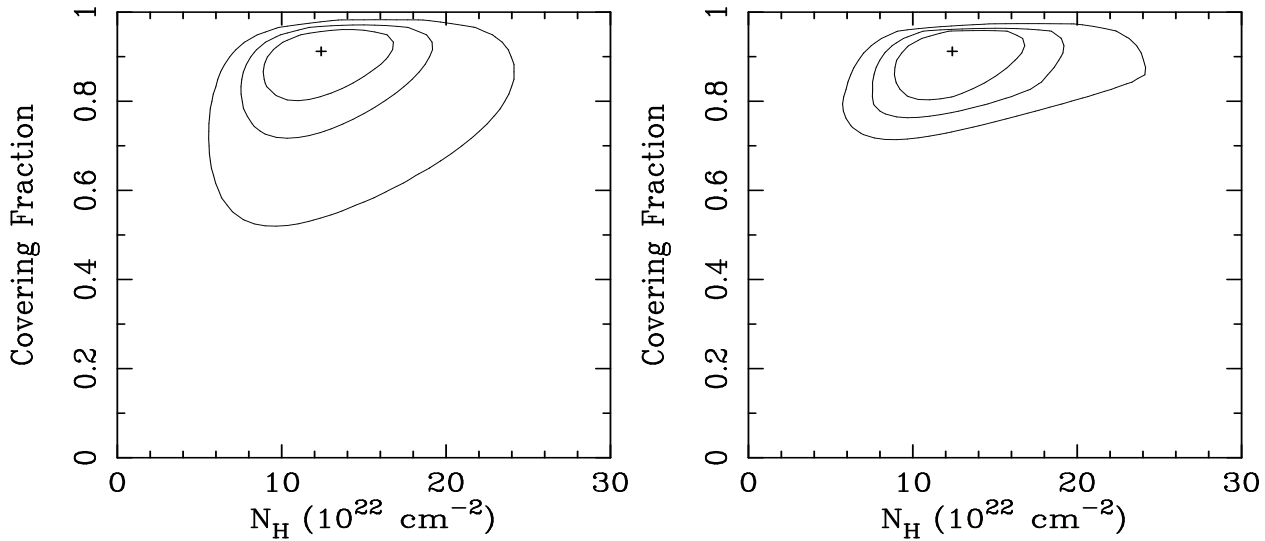

\hbox{
\psfig{figure=figure5a.ps,height=2.8truein,width=3.25truein,angle=-90}
\psfig{figure=figure5b.ps,height=2.8truein,width=3.25truein,angle=-90}
}
\caption{
Confidence contours for the two partial-covering parameters in the spectral analysis
of \pgfif.  The contours are for 68\%, 90\% and 99\% confidence.  In the 
plot on the left, the photon index, $\Gamma$, is allowed to vary freely.  
On the right, 
$\Gamma$ is limited to the range 1.5--2.5 to keep it within the bounds observed
for Seyfert~1 galaxies. Since intrinsic absorption can 
masquerade as a flattening of the power-law slope, a lower limit on $\Gamma$
narrows the range of acceptable covering fractions.
}
\end{figure}
\clearpage
\begin{figure}[t!]
\hbox{
\psfig{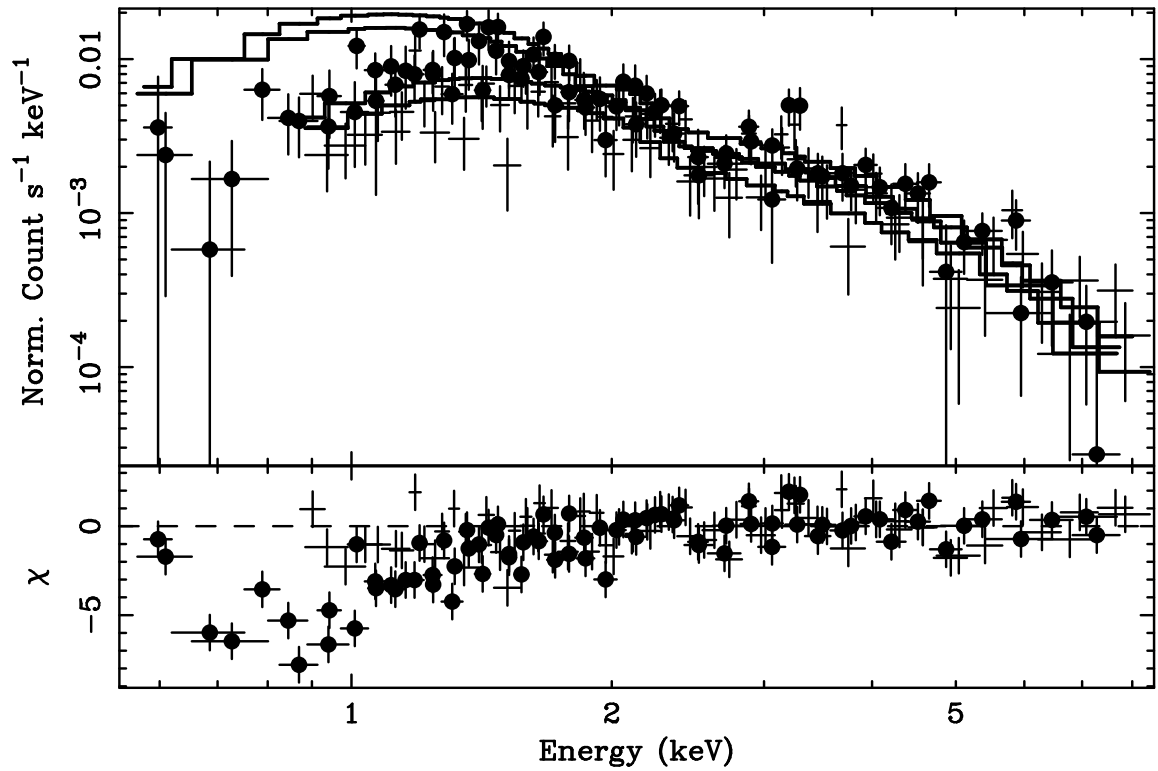}
}
\caption{\asca\ SIS and GIS observed-frame spectra of \pgbal\ fit with a power-law model
above 2~keV which is then extrapolated back to lower energies.  
Filled circles are the data points for the SIS detectors while plain crosses are
for the GIS detectors. 
The ordinate for the lower panel, labeled $\chi$, shows the fit 
residuals in terms of $\sigma$ with error bars of size one.
 Note the low-energy residuals suggestive of
a warm absorber or partial covering model.
}
\end{figure}
\clearpage
\begin{figure}[t!]
\hbox{
\psfig{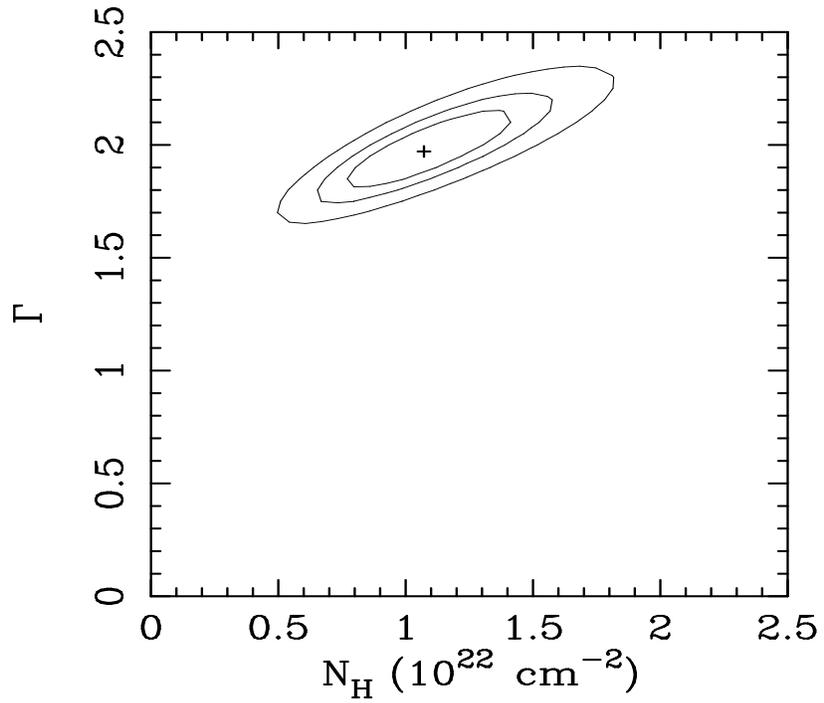}
}
\caption{
Confidence contours in the spectral analysis of the BAL~QSO \pgbal\ 
for the two parameters of interest in the absorbed power-law
model (see model~3 in Table~3).
The contours are for 68\%, 90\% and 99\% confidence.
Note that even within the 99\% confidence contours $\Gamma$ is within the 
range for normal QSOs.
}
\end{figure}
\clearpage
\begin{figure}[t!]
\hbox{
\psfig{figure=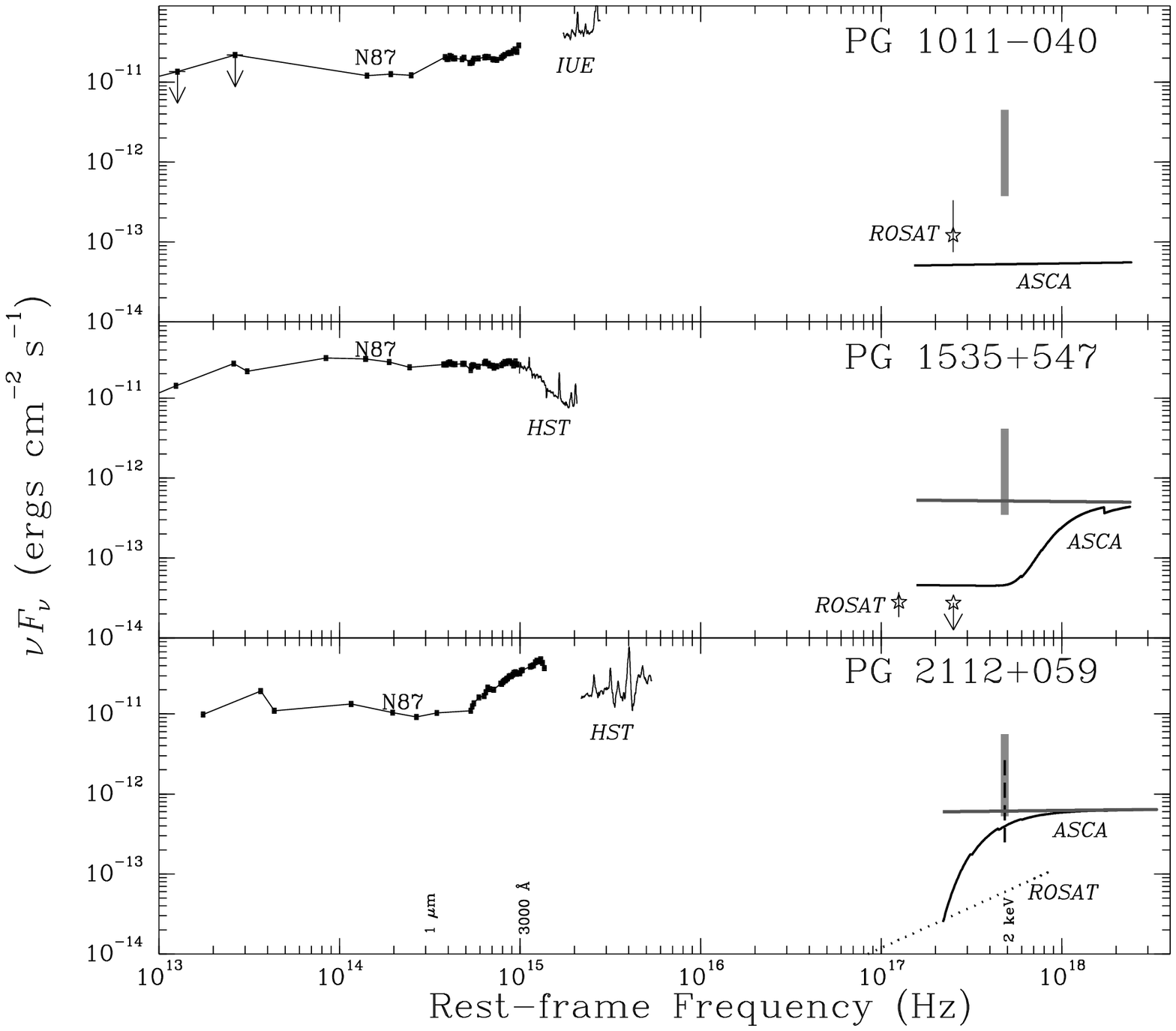,height=5.5truein,width=5.0truein,angle=0}
}
\caption{Spectral energy distributions for each SXW AGN.  Solid squares are infrared and 
optical continuum data from 
Neugebauer et al. (1987; N87), and thin solid curves are UV spectra from \iue\ or 
\hst\ as indicated.  \rosat\ data points are marked with stars for 
\pgten\ and \pgfif, and the \rosat\ best-fitting model for \pgbal\ is 
indicated with a dotted line.  \asca\ data are for the best-fitting
SIS0 model; the thick gray line is the same model corrected for intrinsic absorption
for \pgfif\ and \pgbal.  The vertical gray bar indicates the spread of  
expected 2 keV fluxes from the 3000 \AA\ flux densities of N87 based on the 
typical range of \aox, $-1.7<$\aox\ $<-1.3$; the dashed vertical line for \pgbal\ is
derived from the \hst\ data.
All data have been corrected for Galactic reddening and absorption as described in
the caption to Figure~1. See $\S3$ for further discussion.  
}
\end{figure}
\clearpage
\begin{figure}[t!]
\hbox{
\psfig{figure=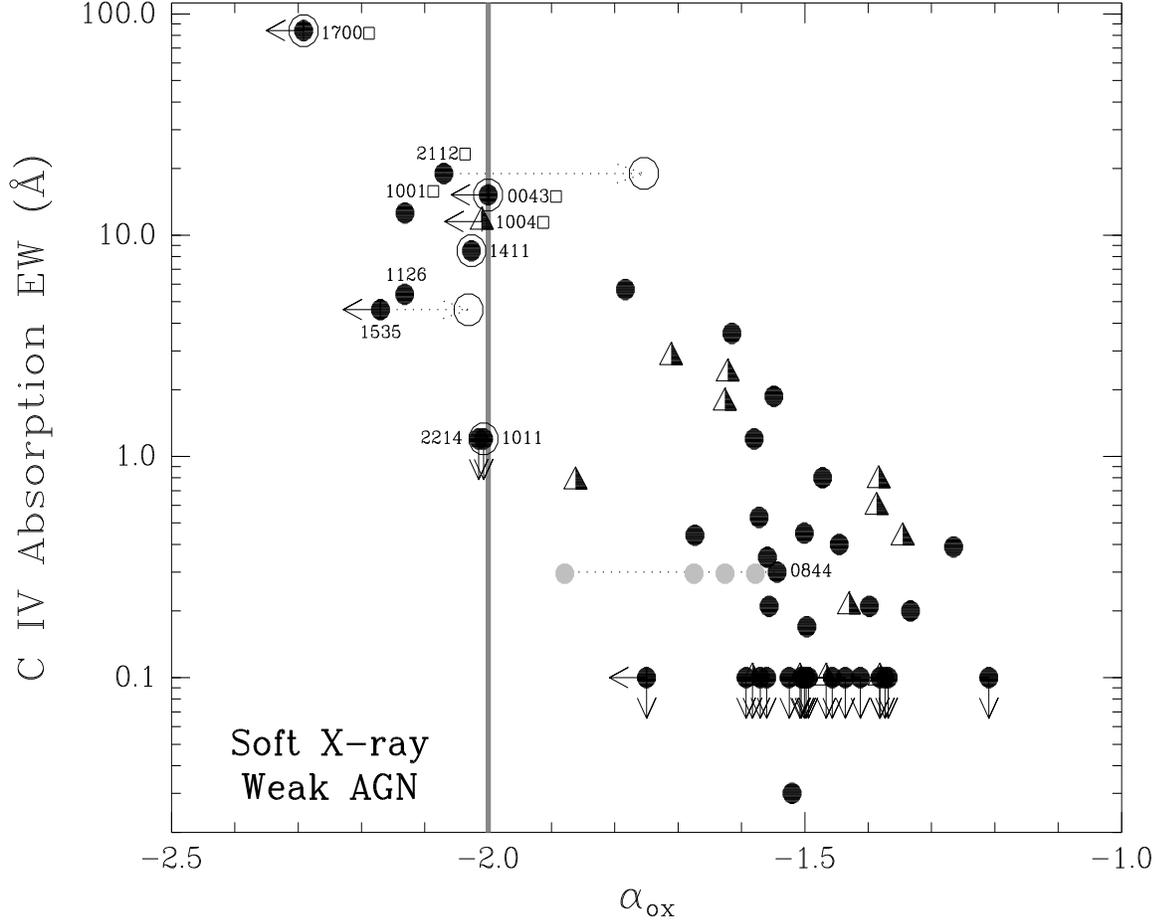,height=4.8truein,width=6.truein,angle=0}
}
\caption{Plot of C~{\sc iv} absorption EW (in the rest frame) versus 
$\alpha_{\rm ox}$ for the BG92 AGN (adapted from BLW).  
Solid dots are radio-quiet AGN and half-filled triangles
are radio-loud AGN.  SXW AGN ($\alpha_{\rm ox}\le-2$) have been labelled with
the right ascension parts of their names.  Large, open circles are SXW AGN with
published 2--10 keV spectral studies.  BAL QSOs are designated with boxes ($\Box$) after
their names. 
PG~0844+349 is also labeled with the hatched circles representing the measured
\aox\ values from four X-ray observations (Wang et al. 2000) other than the 
\rosat\ PSPC one used by BLW.  
Note the remarkable X-ray variability of PG~0844+349, \pgfif, and \pgbal\ as 
indicated with the dotted lines; \pgbal\ is
no longer `soft X-ray weak' according to the definition of BLW. 
}
\end{figure}
\clearpage

\pagebreak[4]
\begin{table}[t!]
\centering
\begin{tabular}{lcccccccc}
\tableline
\tableline
  &     &     &       &   C~{\sc iv}  &  & $N_{\rm H}^{\rm c}$        & SIS/GIS Exp. & Observation \\
Target Name    & $B$ & $z$ & $M_{\rm V}$&EW$^{\rm a}$ (\AA) & $\alpha_{\rm ox}^{\rm b}$ & $(10^{20}$~cm$^{-2})$  & Time$^{\rm d}$ (ks) & Date  \\
\tableline
\pgten                 & 15.5  & 0.058 & $-22.7$ &$<1.2$ & $-2.00$ & 3.78$^1$ & 33.9/38.9  & 1999 Nov 30  \\
\pgfif  & 15.3  & 0.038 & $-22.2$ &4.6&$-2.45$ & 1.35$^1$  & 31.9/37.5 &1999 Jun 27 \\
\pgbal              & 15.5  & 0.457 & $-27.3$ &19&$-2.11$ & 6.26$^2$  &29.7/36.7 & 1999 Oct 30\\
\tableline
\end{tabular}
\\
\caption{General information and observational specifics for the three SXW AGN.
$^{\rm a}$ Rest-frame C~{\sc iv} absorption EW from BLW.
$^{\rm b}$ Estimated from 3000~\AA\ continuum fluxes 
\citep{NeEtal1987} and \rosat\ PSPC count rates (BLW).
$^{\rm c}$ References for Galactic H~{\sc i} 
column densities 
are indicated with numerical superscripts: 
(1) Murphy et al. (1996), (2) Lockman \& Savage (1995).
$^{\rm d}$ All SIS/GIS observations were performed in 1-CCD/pulse-height mode.
}
\end{table}
\nocite{LoSa1995,MuLoLaEl1996}
\clearpage
\begin{table}
\centering
\begin{tabular}{lccccc}
\tableline \tableline
& 
\multicolumn{2}{c}{Count Rate$^{\rm a}$ ($10^{-3}$ ct s$^{-1}$)}& 
\multicolumn{2}{c}{Flux$^{\rm b}$ (10$^{-13}$ erg~cm$^{-2}$~s$^{-1}$)} &
\\
Target Name  & 
SIS0/SIS1 & 
GIS2/GIS3 & 
0.5--2.0 keV & 
2.0--10.0 keV &
$\alpha_{\rm ox}^{\rm c}$ \\

\tableline
\pgten	& $5.7\pm0.7$/$4.7\pm0.6$  & $<1.7$/$2.8\pm0.7$    &$0.7\pm0.3$ &$1.1\pm0.3$& $-2.02$  \\
\pgfif\ & $6.3\pm0.8$/$6.3\pm0.8$  & $6.7\pm0.7$/$9.0\pm0.8$ &$0.59^{+0.12}_{-0.13}$&$4.1\pm0.9$ & $-2.03$  \\
\pgbal  & $21\pm1$/$16\pm1$        & $9.6\pm0.7$/$14\pm0.8$  &$2.4\pm0.2$  &$7.5\pm0.6$ & $-1.75$  \\
\tableline
\end{tabular}
\\
\caption{Observed \asca\ parameters.
$^{\rm a}$ Background-subtracted count rates; 0.6--9.5/0.9--9.5 keV for SIS/GIS. 
$^{\rm b}$ Integrated fluxes are for the best-fitting SIS0 models as described in the text.
Errors represent the range in model fluxes given the 90\% confidence range in the model 
normalization
with all other parameters fixed.
$^{\rm c}$ Calculated from 3000 \AA\ continuum fluxes \citep{NeEtal1987} and the 
best-fitting SIS0 model.
}
\end{table}
\nocite{LoSa1995,MuLoLaEl1996}
\clearpage

\newpage
\begin{table}[t!]
\centering
\begin{tabular}{lclccc}
\tableline \tableline
Model$^{\rm a}$ & $\Gamma$ & Parameter & Values & $\chi^2/\nu$& $P(\chi^2|\nu)^{\rm b}$ \\
\tableline
\multicolumn{6}{c}{\pgten}\\
\multicolumn{6}{l}{1. Power law} \\
	& $1.93^{+0.44}_{-0.39}$     & ...  & ... & 53.6/47 & 0.24 \\
\tableline
\multicolumn{6}{c}{\pgfif}\\
\multicolumn{6}{l}{1. Power law} \\
	& $0.45^{+0.17}_{-0.22}$     & ...  & ... & 134.7/119 & 0.16 \\
\multicolumn{6}{l}{2. Power law with intrinsic partial-covering absorption}\\
	& $2.02^{+0.92}_{-0.95}$	& $N_{\rm H}$ ($10^{22}$~cm$^{-2}$) & $12.3^{+8.3}_{-5.4}$
	& 117.5/117  & 0.47 \\
	&				& Coverage Fraction	& $0.91^{+0.07}_{-0.26}$ & &  \\
\tableline
\multicolumn{6}{c}{\pgbal}\\
\multicolumn{6}{l}{1. Power law} \\

	& $1.44^{+0.08}_{-0.09}$     & ...  & ... & 182.8/151 & 0.04 \\
\multicolumn{6}{l}{2. Power law, $E>3$~keV rest frame (2~keV observed frame)} \\
	
	& $1.94^{+0.23}_{-0.21}$     & ...  & ... & 69.0/76 & 0.70 \\
\multicolumn{6}{l}{3. Power law with intrinsic, neutral absorption}\\
	& $1.97^{+0.26}_{-0.23}$ & $N_{\rm H}$ ($10^{22}$~cm$^{-2}$) & $1.07^{+0.51}_{-0.37}$
	& 142.2/150 & 0.66 \\ 
\multicolumn{6}{l}{4. Power law with intrinsic partial-covering absorption}\\
	& $1.98^{+0.40}_{-0.27}$	& $N_{\rm H}$ ($10^{22}$~cm$^{-2}$) & $1.0^{+1.4}_{-0.49}$
	& 142.0/149  & 0.65 \\
	&				& Coverage Fraction	& $0.97^{+0.03}_{-0.26}$ & &  \\
\multicolumn{6}{l}{5. Power law with an ionized absorption edge}\\
	&$1.88^{+0.28}_{-0.24}$ 	& Rest-frame Edge Energy (keV)	& $0.66^{+0.09}_{-0.15}$
	& 142.7/149  & 0.63 \\
	&				& Maximum Optical Depth, $\tau$ & $9.7^{+>10}_{-7.9}$ & &  \\
\tableline
\end{tabular}
\caption{
Parameters of model fits to the \asca\ data for all three sources.
$^{\rm a}$All models have fixed Galactic absorption as given in Table~1.  
Models with intrinsic absorption have $z$ fixed and assume neutral gas with 
solar abundances unless otherwise noted.
$^{\rm b}$ The probability, if the given model were correct, that this value of $\chi^2/\nu$ 
or greater would be obtained where $\nu$ is the number of degrees of freedom.
}
\end{table}
\clearpage

\end{document}